\documentclass[12pt]{iopart}

\usepackage{graphicx}
\usepackage[utf8]{inputenc}

\newcommand{\Pu}{$^{241}\textrm{Pu}$~}
\newcommand{\T}{$^{3}\textrm{H}$~}
\newcommand{\Tm}{$^{171}\textrm{Tm}$~}
\newcommand{\Sm}{$^{151}\textrm{Sm}$~}
\newcommand{\Ni}{$^{63}\textrm{Ni}$~}

\begin{document}

\title{Plutonium-241 as a possible isotope for neutrino mass measurement and capture}

\author{Nicolo de Groot}

\address{ Institute for Mathematics, Astrophysics and Particle Physics (IMAPP), Radboud University Nijmegen, Heyendaalseweg 135,  6525 AJ Nijmegen, The Netherlands}
\ead{N.deGroot@science.ru.nl}



\begin{abstract}
Tritium has been the isotope of choice for measurements of the neutrino mass
and planned detection of the relic neutrino background. The low mass of \T
leads to large recoil energy of the nucleus. This  has emerged as limiting factor 
for both measurements. We investigate \Pu as an alternative.  
The recoil is 80 times smaller and it has a similar decay energy and lifetime as {\T}.  
We evaluate for the first time its soft-neutrino capture cross-section 
and find $(\sigma v)_{\nu} = 1.52 \times 10^{-45}$cm$^2$.  This is 40\% of the capture 
cross-section for tritium and makes \Pu an interesting alternative 
for \T
\end{abstract}
\submitto{\jpg}

%

\section{Introduction}
\label{sec:intro}
The measurement of the absolute neutrino mass and the detection of the cosmic neutrino background (CNB)
are two of the most exciting prospects in neutrino physics for the next two decades.
They are closely related.  The mass measurement is looking for a deformation of the end of the beta decay spectrum  caused by the finite neutrino mass, while the CNB search is looking for a small peak from neutrino
capture which is separated by twice the lightest neutrino mass from the end point. Both measurements
require an experimental precision and a theoretical understanding of the end of the beta decay spectrum 
at the level of the neutrino mass, which is of the order of tens of meV.

Candidate nuclides for the mass measurement should first of all have moderate beta-decay energy. Already for isotopes with a decay energy of 10-100 keV, an experimental precision of 1ppm would be required. 
A lifetime of 1-20 years is optimal since it leads to a reasonable event rate and relatively stable measurement 
conditions. Rhenium-187 has the lowest energy of the known beta-decaying isotopes at 2.6 keV, but due to its lifetime of 60 Gyr there would be no events close to the endpoint within a human lifetime.
In addition the daughter isotope needs to be stable on the timescale of the experiment. Ruthenium-106 
for example, does have a significant neutrino capture cross-section, a decay energy of 39.4 keV and a half-life of 1.48 years,
but the daughter isotope Rhodium-106 decays within 30 seconds with a decay energy of several MeV and the
emission of several gammas, making a clean measurement of the endpoint impossible.

For candidate nuclides for detection of the CNB  there is an additional requirement to have a large neutrino capture cross-section to get an acceptable even rate. The capture cross-section has been evaluated
for a large number of isotopes~\cite{Cocco2008}. From this study \T, which has always been the isotope of choice for the mass measurement,
is the isotope that best matches the requirements. 
 The KATRIN experiment is using molecular tritium in its current world best neutrino mass limit~\cite{katrin2022}. Their current measurement is still dominated by its statistical error, but with more data it will be limited by the energy spread of the vibrational and rotational excitation spectrum of the daughter molecule $(^3\textrm{H}^3\textrm{He})^+$. The decaying tritium nucleus recoils against the electron and picks up 3.4 eV kinetic energy due to its relatively low mass. This is enough to bring the daughter molecule in one of several of the excited states and leads to an energy spread of 0.36 eV.
 This is the reason that future experiments, like Project8~\cite{Project8-2017} are looking at atomic tritium.
The PTOLEMY experiment~\cite{Ptolemy2019} also plans to use atomic \T loaded on graphene to measure the 
mass and in a later stage observe the CNB.

A recent study~\cite{Cheipesh2021} found that the zero-point motion of the \T bound to the surface of graphene,
or for that matter any surface, leads to an energy spread of $\Delta E \sim 0.5$ eV. 
They advocate the investigation of nuclei with $A>100$ and define a figure of merit for this effect:
\begin{equation}
\gamma = (Q^2 m_e / m_{\textrm{nucl}}^3)^{1/4}
\end{equation}
and propose \Tm and \Sm as candidate isotopes, which have a $\gamma$ which is an order of magnitude 
lower than for \T.

\section{Beta decay of \Pu}
Plutonium-241 is an isotope which is created by double neutron capture on Plutonium-239.
About 12\% of the plutonium in spent nuclear fuel is \Pu.
It mainly undergoes beta decay to $^{241}\textrm{Am}$ with a decay energy of 20.78 keV and a 
half life of 14.4 year, numbers very similar to those of \T and one would expect the experimental setup for 
atomic \T to work for \Pu as well. What makes \Pu interesting for the mass measurement is the large
mass of the isotope and as a consequence its low recoil of 47 meV. The $\gamma$ factor is 25 times
smaller than for $^3$H, which according to~\cite{Cheipesh2021} should allow for observation of the CNB
for neutrino masses of $m_{\nu} > 30 \textrm{meV}$.

The daughter nuclide $^{241}\textrm{Am}$ decays trough $\alpha$ decay to $^{237}$Np with a half life
of 432 years. $^{237}$Np itself has a half life of 2 million  years and can be considered stable.  In
addition \Pu has a  small ($2.4 \cdot 10^{-5}$) probability to have an $\alpha$ decay to $^{237}$U.
 The energies of the $\alpha$ particles, and the subsequent $\gamma$'s are well outside the relevant energy
 window around the endpoint.  The biggest challenge comes from the $^{237}$U. It has several $\beta$
 decays to $^{237}$Np with a half life of 6.7 days and an energy up to 459 keV.  The total rate from
 the $^{237}$U is about five orders of magnitude smaller than for \Pu, but near the endpoint the \Pu
 spectrum is falling rapidly and $^{237}$U is still on a plateau.  At around 20 eV before the endpoint
 the rates become similar and a mass measurement becomes impossible without special measures 
 to remove the $^{237}$U on a timescale much shorter than its lifetime. There are two possible
 ways to reduce the $^{237}$U background. First the $\alpha$ decay to $^{237}$U  causes a 
 considerable recoil of around 150 keV. In the design of the target this could conceivably used to
 remove part of the  $^{237}$U from the active target area.  Secondly, the $\beta$ decay to $^{237}$Np is
  never to the ground state and always followed by  a $\gamma$ decay which can be used as a veto.

\section{Neutrino capture cross-section on \Pu}
In order to calculate the neutrino capture cross-section for \Pu, we follow a procedure as developed
in~\cite{Mikulenko2021}, which we will summarize briefly below.
We consider the two related weak processes:
\begin{eqnarray}
^A_{Z}X & \rightarrow & ^A_{Z+1}Y + e^- + \bar{\nu_e}\nonumber \\
\nu + ^A_{Z}X & \rightarrow & ^A_{Z+1}Y + e^-
\end{eqnarray}

Using Fermi's Golden Rule, we can write the differential beta-decay rate $d\Gamma_{\beta}$ and the neutrino capture cross-section as:

\begin{eqnarray}
d\Gamma_{\beta} &=& \frac{1}{2\pi^3} \times p_{\nu} E_{\nu} E_e dE_e  \times W_{\beta}(p_e, p_{\nu}) \nonumber \\
(\sigma v)_{\nu} &=& \lim_{p_{\nu} \rightarrow 0} \frac{1}{\pi} \times p_e E_e \times W_{\nu}(p_e, p_{\nu}).
\end{eqnarray}

Here $W_{\beta}(p_e, p_{\nu}$) is the average transition rate for the decay of an atom emitting two leptons in a plane wave with momenta $p_e$ and $p_{\nu}$ and  $W_{\nu}(p_e, p_{\nu})$ the 
average transition rate for the capture of a neutrino with momentum $p_{\nu}$ with the emission
of an electron with $p_e$. They are obtained by integrating the transition amplitudes squared over 
the directions of the leptons, summing over the quantum numbers of the outgoing particles, and averaging over the incoming particles.

For the CNB neutrinos, with $p_{\nu}<<m_{\nu}$ we have:
\begin{equation}
W_{\nu}(p_e, 0) = \frac{1}{2} \lim_{p_{\nu} \rightarrow 0} W_{\beta}(p_e, p_{\nu})
\end{equation}

Earlier calculations of the neutrino cross-sections ~\cite{Cocco2008} have been performed 
for two kinds of beta decays. Allowed transitions are decays where the parent and daughter isotope have the same quantum numbers and no angular momentum is carried away by the lepton pair. In 
this case the transition amplitudes can be approximated by a constant and their ratio can be taken 
as 1, after which it is straightforward to express the capture cross-section in terms of the total decay width of the isotope. This is not true for forbidden decays where the quantum numbers are different for 
mother and daughter and there will be a dependency on the momenta.

For a unique forbidden transition, there is only one term contributing. The matrix element and the kinematic factor factorize and the ratio 
of the transition rates can be calculated after which the neutrino capture cross-section is again expressed as function of the total width.
For non-unique forbidden transition the matrix elements contains several terms with each its own dependency on the momenta and such a
calculation can not be performed. 

The novel  approach of~\cite{Mikulenko2021} is using the fact that $W_{\beta}(p_e, p_{\nu})$ is an analytical function of the momenta, 
and evaluating the differential decay rate for the same kinematic region as for neutrino capture. For small enough values of $p_{\nu}$ a linear 
approximation can be used and $(\sigma v)_{\nu}$ can be found by extrapolating to $p_{\nu} = 0$:

\begin{equation}
(\sigma v)_{\nu} [1 + \alpha_1 p_{\nu}/Q + O(p_\nu^2/Q^2)] = 
\frac{\hbar^3 c^2 \pi^2}{p_{\nu}^2} \frac{d\Gamma_{\beta}}{dE_e} 
\label{eq:sigma}
\end{equation}
Where we put the factors $\hbar$ and $c$ back in. This procedure requires a well-measured decay spectrum.

\section{Results}
For the extraction of $(\sigma v)_{\nu}$ from ~\ref{eq:sigma} the beta spectrum of \Pu has been measured very precisely\cite{Loidl2010}. We use a parametrization of this data as our main set.
 As a cross-check data set we generate a spectrum using BetaShape~\cite{Mougeot-betashape, Mougeot2014} using an energy bin width of 0.06 eV. It is useful to point out that the measured
 spectrum has been used as one of the benchmark processes for the BetaShape program.

\begin{figure}[ht]
     \centering
   \includegraphics[width=0.8\textwidth]{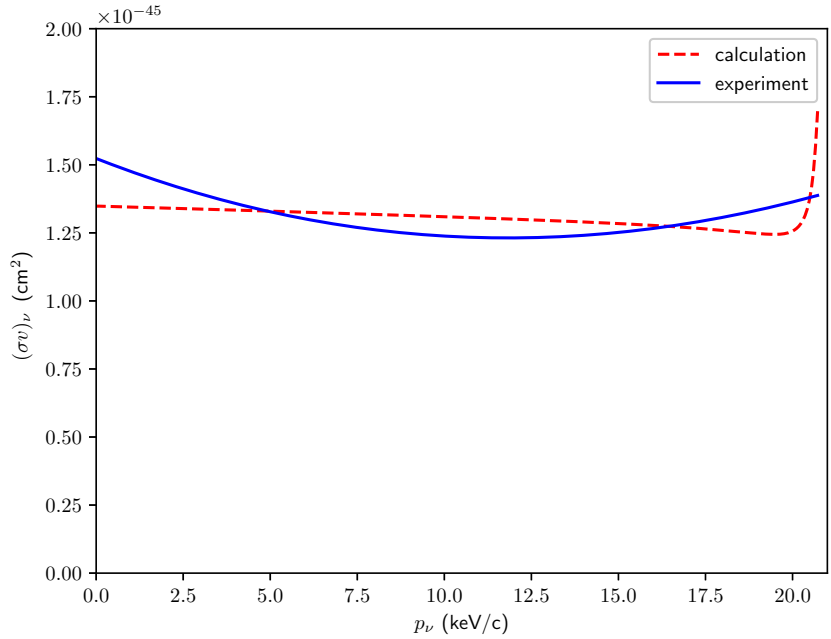}
     \caption{Calculated neutrino caption cross-section for \Pu as a function of the neutrino momentum
     in the $\beta$ decay}
     \label{fig:sigma}
\end{figure}

 The results are shown in figure~\ref{fig:sigma}.
We see that below 5 keV, the experimental data is well described by a linear function and we find
$(\sigma v)_{\nu} = 1.52 \cdot 10^{-45}\,\textrm{cm}^2$ and $\alpha_1 = -0.51$. The calculated spectrum gives
a value of $(\sigma v)_{\nu} = 1.34 \cdot 10^{-45}\,\textrm{cm}^2$, which is fairly close. 

As a cross-check we also evaluate the results for the two isotopes \Sm and \Tm. We generated
both spectra with BetaShape using energy bins of 0.2 eV and fitted the spectrum, where we reproduced
the values found in~\cite{Mikulenko2021}, which were confirmed in a similar analysis~\cite{Brdar2022}.
In addition we performed the same procedure for the \Ni
isotope and recovered the calculated cross-section from~\cite{Cocco2008}.

\section{Conclusion}

\begin{table}[!htb]
\caption{\label{tab1}Neutrino capture cross-sections for different isotopes. The values for \T and \Ni 
are taken from~\cite{Cocco2008}.}
  \begin{indented}
  \lineup
  \item[] \begin{tabular}{@{}lllll}
    \br
     Isotope   &  $Q$ (keV) & $t_{1/2}$ (yr) & $(\sigma v)_{\nu} (10^{-46}\textrm{cm}^2$) & $\gamma/\gamma_{^3H}$ \\
  \mr
     \T   & 18.6   &  \012.3  &  39.2  & 1.0 \\
     \Ni &   66.9     &  100   &\00.069  &  0.19\\
     \Sm & 76.6  &  \090 &  \00.048 & 0.10 \\
     \Tm & 96.5  &  \0\01.92 & \01.2 & 0.11 \\
     \Pu &  20.8 &  \014.4 & 15.2 & 0.039 \\
   \br
       \end{tabular}
\end{indented}
\end{table}

We have for the first time estimated the neutrino capture cross-section on \Pu and found it
to be $1.52 \cdot 10^{-45}$. The result from the actual beta spectrum and the BetaShape 
calculation agree within 10\%, which gives confidence in this result. The relevant parameters for 
\Pu and other candidate isotopes are shown in table~\ref{tab1}. If the energy uncertainty for
\T cannot be solved, \Pu seems to be a promising replacement for at least the neutrino 
mass measurement, provided the $^{237}$U can be removed or its decay vetoed. It has an energy and 
life time which is very similar to \T, and an experiment designed for \T will also work for \Pu. This is 
not the case for \Sm or \Tm which have a substantial larger energy. The energy uncertainty is more
than twice smaller than for these and would according to~\cite{Cheipesh2021}  allow for a 
CNB observation for neutrinos with $m_{\nu}>30\textrm{meV}$. The expected rate is lower than for \T 
but at least 10 times higher than for \Tm and the calculation is based on an actual spectrum.

\ack
The author is  grateful to Oleksii Mikulenko of Institute Lorentz in Leiden, Geon-Bo Kim of Lawrence 
Livermore National Laboratory and Adriaan K\"onig  of Radboud University for their valuable comments.\\



\bibliography{Pu241}





\end{document}